\documentclass[11pt]{article}
\usepackage{jheppub}
\usepackage[utf8]{inputenc}
\usepackage{amsmath}
\usepackage{tikz}
\usepackage{tkz-euclide}
\usepackage{dsfont}
\usepackage{amssymb}
\usepackage{amsmath}
\usepackage{mathrsfs}
\usepackage{appendix}
\usepackage{bbold}
\usepackage{color}

\def\ie{\begin{equation}\begin{aligned}}
\def\fe{\end{aligned}\end{equation}}
\begin{document}

\title{Holographic Consequences of Heterotic String Theory beyond its Supergravity Approximation  } 

 \author[1,10]{Leon Berdichevsky Acosta,}
 \author[2]{Salman Sajad Wani,}
 \author[1]{Arshid Shabir,}
\author[4]{Abrar Ahmed Naqash,}
\author[5]{Imtiyaz Ahmad Bhat,}
\author[6]{Naveed Ahmad Shah,}
\author[7]{Hrishikesh Patel,}
\author[2]{Saif Al-Kuwari,}
\author[1,3,8,9]{Mir Faizal}

\affiliation[1]{Canadian Quantum Research Center 204-3002, 32 Ave Vernon, BC V1T 2L7 Canada}
\affiliation[2]{Qatar Center for Quantum Computing, Hamad Bin Khalifa University, Doha, Qatar}
\affiliation[3]{Irving K. Barber School of Arts and Sciences, University of British Columbia Okanagan, Kelowna, BC V1V 1V7, Canada}
\affiliation[4]{Department of Physics, National Institute of Technology, Srinagar, Srinagar, Jammu and Kashmir 190006, India}
\affiliation[5]{Department of Physics, Central University of Kashmir, Ganderbal, Jammu and  Kashmir, 191201 India}
\affiliation[6]{Department of Physics, Aligarh Muslim University, Aligarh, Uttar Pradesh, 202002  India}
\affiliation[7]{Department of Physics and Astronomy, and Quantum Matter Institute,
University of British Columbia, Vancouver, BC V6T 1Z1, Canada }
\affiliation[8]{Department of Mathematical Sciences, Durham University, Upper Mountjoy, Stockton Road, Durham DH1 3LE, UK}
\affiliation[9]{Faculty of Sciences, Hasselt University, Agoralaan Gebouw D, Diepenbeek, 3590 Belgium}
\affiliation[10]{Department of Actuarial Science and Insurance,
ITAM, Mexico City,  01080 Mexico }

\emailAdd{leonb@itam.mx}
\emailAdd{sawa54992@hbku.edu.qa}
\emailAdd{aslone186@gmail.com}
\emailAdd{abrar.naqash1234@gmail.com}
\emailAdd{immy.salam@gmail.com}
\emailAdd{shahnaveed75@gmail.com}
\emailAdd{hrishikeshnitin@gmail.com}
\emailAdd{smalkuwari@hbku.edu.qa}
\emailAdd{mirfaizalmir@gmail.com}

\abstract{In this work, we study the effects of stringy corrections on the low-energy effective action derived from heterotic string theory beyond its supergravity approximation. Compactifying the ten-dimensional theory with these stringy corrections produces an effective action for a scalar field whose higher-derivative term is governed by a single coefficient that depends on the internal volume, average curvature, and flux of the compactification manifold.
The higher-derivative coupling imported from compactification shifts the Breitenlohner-Freedman stability bound by an amount set by the relative strengths of internal flux and curvature, relaxing it in flux-dominated vacua and tightening it in curvature-dominated ones. Furthermore, we analyze how the stringy corrections shift the scaling dimensions of the dual operators, track the resulting renormalization-group flow, and investigate the higher-derivative term using a holographic Lee–Wick regulator. 
In holographic superconductors, stringy corrections lower the effective bulk mass and raise the critical temperature when flux dominates, but have the opposite effect when curvature dominates. }

\maketitle
\section{Introduction}
 The supergravity theories describe the low-energy effective field-theory approximation to string theory \cite{super}. In this approximation, the theory can be described as a standard field theory, and any purely stringy effect is completely suppressed.
Thus, in the supergravity approximation, the effects produced due to the finite length of strings \(l_s\) can be neglected \cite{super}. Now in this regime, as the stringy structures are suppressed, ordinary field-theoretic degrees of freedom suffice. To uncover intrinsically stringy effects, one must go beyond this approximation and include the higher-derivative operators generated by the finite string length \(l_s^{2}\). These terms—called \(\alpha'\)-corrections because \(\alpha' = l_s^{2}\)—encode the leading imprint of the underlying stringy dynamics on the low-energy effective action and give rise to phenomena absent in pure supergravity.

It is possible to use string scattering amplitudes to obtain \(\alpha^{\prime}\) corrections to the supergravity action \cite{GrossSloan1987, 2, Metsaev87}. These corrections have also been obtained using \(\beta\)-function calculations \cite{1a, 2a, 4a, 5a}. In this approach, it has been demonstrated that the vanishing of the higher-loop string \(\beta\)-function is equivalent to higher-derivative \(\alpha^{\prime}\) corrections to the supergravity actions. 
Such higher-derivative terms are also important in moduli stabilization \cite{m1, m2}. Such higher-derivative \(\alpha^{\prime}\) corrections to F-theory have also been used to obtain the higher-derivative \(\alpha^{\prime}\) corrections to type IIB supergravity \cite{f1}. It is possible to obtain a non-trivial vacuum profile for the four-dimensional effective action from the compactification on elliptically fibered Calabi–Yau fourfolds \cite{f2}. 
The eleven-dimensional supergravity corrected by higher-derivative terms has been used to investigate higher-derivative corrections to the M5-brane action \cite{m4}. The higher-derivative corrections to the AdS supergravity actions have been used to investigate the geometry of \(N\) M5-branes \cite{m41}. 
The Kaluza–Klein reduction of \(\alpha^{\prime}\)-corrected type IIB supergravity has been done on a Calabi–Yau threefold by considering infinitesimal Kähler deformations of the Calabi–Yau background \cite{m5}. Hence, the four-dimensional supergravity contains the four-derivative couplings of the Kähler moduli fields. Thus, it is important to consider the effects of higher-derivative corrections to the supergravity actions.  

The higher-derivative \(\alpha^{\prime}\) corrections to the supergravity action for type IIB string theory have been obtained by investigating the invariance of such low-energy effective actions under gauge symmetry and the T-duality constraints \cite{td12}. The NS-NS couplings have been found in this scheme and expressed in terms of the Riemann curvature, the field strength of the two-form field, and its derivative using field redefinition. 
Such higher-derivative corrections to the low-energy effective action have also been obtained by imposing the gauge symmetries and T-duality on the effective action for R-R coupling \cite{td14}. The \(\alpha^{\prime}\) corrections to the DBI action describing D-branes have been investigated using T-duality constraints \cite{td15}. The \(\alpha^{\prime}\) corrections to the effective action of an O-plane can also be studied using T-duality \cite{td16}.   

Although the AdS/CFT correspondence is most often studied using supergravity solutions, it is conjectured to be exact in full string theory \cite{q}. It therefore extends well beyond the supergravity approximation.  
Since the correspondence holds for the complete string theory on AdS \cite{q}, one can systematically investigate stringy \(\alpha^{\prime}\) corrections to the duality. 
Thus, the CFT duals of these stringy \(\alpha^{\prime}\) corrections to supergravity can also be studied using the AdS/CFT correspondence \cite{q1,q2}.
Such stringy higher-derivative corrections in the AdS bulk correspond to \(1/N\) corrections in the boundary CFT, where \(N\) is the number of colors in the dual gauge theory \cite{n1,n2}. In fact, the effect of stringy higher-derivative corrections on the holographic dictionary between AdS\(_2\) and CFT\(_1\) has also been investigated \cite{cft1}. In this study, the holographic dual to the Wald entropy of the BPS black hole was obtained from an operator dual to a composite bulk field. Thus, it is important to study the CFT dual to higher-derivative \(\alpha^{\prime}\) corrections in the bulk. The \(\alpha^{\prime}\)-corrected type IIB supergravity has been used to investigate the effect of such higher-derivative corrections on the quantum chaos of the boundary field theory \cite{cft2}. The higher-derivative corrections to the type IIB supergravity on AdS\(_5 \times\)S\(^5/Z_n\) have been used to obtain \(1/N\) corrections to the holographic Weyl anomaly \cite{cft4}. The stringy higher-derivative corrections to the ratio of shear viscosity to entropy density have been holographically investigated using the AdS/CFT correspondence \cite{cft5}. It has been observed that these stringy higher-derivative terms produce \(1/N\) corrections, and this causes a violation of the bound for the shear-viscosity-to-entropy ratio. It has also been observed that turning on the chemical potential would lead to an even larger violation of the bound. 

In this paper, we analyze the holographic dual to higher-derivative corrections in the AdS. To simplify our analysis, we only consider the holographic dual to the higher-derivative corrections of a scalar field in AdS. This is obtained as an effective action using the stringy corrections to heterotic string theory beyond its supergravity approximation \cite{5a,GrossSloan1987}. 
We analyze the effect of higher-derivative corrections on the Breitenlohner–Freedman (BF) bound. Such corrections to the BF bound have been discussed in the context of AdS/CFT corrections beyond its supergravity approximation \cite{q1, q2}. We will explicitly investigate how the compactification of the \(\alpha^\prime\)-corrected effective action modifies the dual CFT. It has been suggested that the BF bound will be modified due to stringy higher-derivative corrections, and this can have important consequences for the holographic model of color superconductivity \cite{bf1}. The modification of the BF bound by higher-derivative corrections has also been used to construct a novel holographic superconductor \cite{bf2}. The analogue of the BF bound for a non-relativistic theory has also been investigated \cite{bf5}. The holographic superconductors whose dual black holes have scalar hair of mass near the BF bound have been constructed \cite{bf6}. This has been done using low temperatures in the probe limit. Thus, it is important to study the effects of higher-derivative corrections on the BF bound of a CFT. We perform a rigorous numerical study of the BF bound for a CFT dual to such higher-derivative terms in AdS. We also analyze the effects of various parameters of the model on the BF bound.  

One important motivation for studying \(\alpha^{\prime}\) corrections to supergravity is the solution to the problem of moduli stabilization \cite{10az}. Moduli stabilization can be accomplished in any of the three superstring theory compactifications by different mechanisms. In type IIB, the flat directions of the moduli space can be lifted by turning on fluxes and by the leading \(\alpha^{\prime}\) correction to the WZ action. Such corrections occur at the order of \(\alpha^{\prime 2}\) and have been extensively studied due to their importance in the anomaly-cancellation mechanism \cite{11az,12az,13az}. In the case of supersymmetric compactifications of the heterotic string, the stabilization of all moduli can be accomplished by fractional fluxes \cite{gc0}, non-perturbative contributions to the effective action from gaugino condensation \cite{gc1}, higher-order perturbative corrections in \(g_s\) \cite{gc3}, turning on three-form flux in combination with hidden-sector gaugino condensation \cite{gc4} as well as by \(\alpha^{\prime}\)-corrections to the K\"ahler potential \cite{gc5}.

We observe that higher-derivative corrections to the effective action of the bulk fields in both type IIA and type IIB superstring theories start only at order \(\alpha^{\prime 3}\) \cite{ap1}. On the other hand, the higher-derivative corrections to the bulk fields in the heterotic string start  at order \(\alpha^{\prime}\) \cite{37az,GrossSloan1987}. Thus an   advantage of the heterotic set-up is that its first stringy modification to supergravity appears already at linear order in $\alpha'$. 
Using heterotic string theory, we will analyze the effect of these corrections on the effective field theory of a scalar field.
After the warped compactification, the bulk scalar sector acquires a single higher-derivative contribution proportional to $\Box^{2}\phi$, whose overall coefficient is fixed solely by the internal volume, average curvature, and flux.  Since there is only one such operator, its effect on AdS$_{d+1}$ propagators can be analyzed analytically.  On the CFT$_{d}$ side this translates into a clean $1/N$ correction to two-point functions and operator dimensions, without  the multitude of competing {quartic-curvature} terms and flux–curvature mixings that first appear in type-II theories at order $\alpha'^{3}$.  The heterotic framework therefore provides the most economical example for isolating and understanding stringy higher-derivative effects within the AdS/CFT correspondence.

The massless bosonic fields of the heterotic string are the metric, the Neveu–Schwarz antisymmetric field and the dilaton. In order to satisfy coordinate invariance and the gauge invariance of the form fields, the Lagrangian must be constructed from these fields. Consider the ten-dimensional heterotic supergravity corrected to first order in \(\alpha^{\prime}\) by all the terms involving the bulk fields and consider the warped compactification of the ten-dimensional spacetime on a product of a \((d+1)\)-dimensional AdS space and a \((9-d)\)-dimensional compact space \(M\). Such compactification gives rise to an infinite tower of massive Kaluza–Klein scalar fields. Stability of solutions on AdS\(_{d+1}\) imposes a lower limit on the mass of scalar fields, known as the BF bound. According to the AdS\(_{d+1}\)/CFT\(_d\) dictionary, scalar fields on AdS\(_{d+1}\) are dual to scalar conformal operators of the boundary CFT\(_d\), and the masses of the scalar fields determine the conformal dimension of such operators. As a consequence, the BF bound on the mass implies that the associated conformal dimension is always a real number.

\section{Effective Action}
The supergravity action emerging from string theory is known to receive higher derivative $\alpha^{\prime}$ corrections. These corrections, which originate from loop effects and higher string modes in the full theory, modify the low–energy dynamics of the bulk gravitational theory. In the context of the AdS/CFT correspondence, the complete string theory on an AdS background, including all $\alpha^{\prime}$ corrections, remains dual to a conformal field theory (CFT). In this context, the $\alpha^{\prime}$ corrections in the bulk map onto subleading $1/N$ corrections in the boundary CFT \cite{q, n1, n2}. In fact, significant effort has been devoted to investigating the CFT dual to an $\alpha^{\prime}$ corrected supergravity action \cite{q1,q2}, and one finds that these corrections further lead to a modification of the  BF  bound into a generalized BF bound. This generalized BF bound encapsulates the modifications in the stability criteria both of the bulk theory and, through the AdS/CFT dictionary, of the unitarity and the scaling dimensions of operators in the dual CFT.
In heterotic string theory, the ten dimensional effective action receives higher derivative $\alpha^{\prime}$ corrections.

Starting from the heterotic world–sheet $\sigma$-model, one-loop conformal invariance fixes the $\beta$-functions, which, when matched to a spacetime action, yield the $\alpha'$-corrected heterotic supergravity \cite{5a,GrossSloan1987}.
This corrected supergravity action includes the dilaton–curvature mixing due to  the Green–Schwarz mechanism.  
After transforming the original string-frame metric \(g^{S}_{MN}\) to the Einstein-frame metric \(g_{MN}=e^{-\Phi/2}g^{S}_{MN}\) and performing the allowable field redefinitions, these mixed terms can be expressed as a single higher-derivative operator for the canonically normalized dilaton fluctuation \(\phi\).  Here, the dilaton field $\Phi$ is expressed as the constant background  $\langle\Phi\rangle$ plus a fluctuation $\phi$ around that background, \( \Phi =  \langle\Phi\rangle + \phi   \).  
In the supergravity limit, with all \(\alpha^{\prime}\) corrections neglected, the effective action in the Einstein frame for the scalar field \(\phi\) contains the canonical kinetic term.
 However, this action is corrected by stringy effects  beyond the supergravity approximation.
Schematically, in the Einstein frame (after suitable field redefinitions), 
these stringy corrections beyond the supergravity approximation correct the action for $\phi$ by adding higher-derivative terms of the form \cite{5a,GrossSloan1987}
\begin{equation}
S^{(10)}_{\rm HD}\;\supset\;
\frac{\alpha^{\prime}}{2\kappa_{10}^{2}}
\int d^{10}x\,\sqrt{-g_{10}}\;\mathcal{L}_{\rm HD}\,,
\end{equation}
where $\mathcal{L}_{\rm HD}$ is
\begin{equation}
\mathcal{L}_{\rm HD}\;\supset\;
\bigl[1+\gamma_{1}\alpha^{\prime}R_{10}
      +\gamma_{2}\alpha^{\prime}H_{MNP}H^{MNP}+\cdots\bigr]
(\Box_{10}\phi)^{2}\,.
\end{equation}
Here, \(R_{10}\) denotes the ten dimensional Ricci scalar, \(H_{MNP}\) is the NS-NS three form field strength (which encodes the flux degrees of freedom), and the constants \(\gamma_1,\gamma_2,\ldots\) are determined by the details of the string theory and anomaly cancellation through the Green-Schwarz mechanism).  The heterotic action also contained the quadratic-curvature Gauss–Bonnet density; in ten dimensions, this invariant was not topological and therefore had to be retained in the full gravitational sector.
However, after the field redefinitions that isolated the single scalar operator \((\Box\phi)^{2}\), the Gauss–Bonnet density no longer mixed with the dilaton at quadratic order.  Since we focused only on the stringy \((\Box_{10}\phi)^2\) corrections to the effective action for \(\phi\), we neglected the contribution of the Gauss–Bonnet term.

To obtain the effective action in \((d+1)\) dimensions, one assumes that the ten dimensional spacetime factorizes as
\begin{equation}
ds_{10}^2 = g_{MN}^{(10)}dx^Mdx^N = g_{\mu\nu}^{(d+1)}(x)dx^\mu dx^\nu + g_{mn}^{(\mathrm{int})}(y)dy^m dy^n\,,
\end{equation}
so that the determinant factorizes as
\(
\sqrt{-g_{10}} = \sqrt{-g_{(d+1)}(x)}\,\sqrt{g_M(y)}\,,
 \)
and the internal volume is defined by
\begin{equation}
V_M = \int_M d^{9-d}y\,\sqrt{g_M(y)}\,.
\end{equation}
If we further assume that the scalar field is independent of the internal coordinates, i.e.,
 \(
\phi(x,y)=\phi(x)\,,
 \)
then the ten dimensional d'Alembertian reduces to the \((d+1)\) dimensional operator,
 \(
\Box_{10}\phi(x) = \Box_{(d+1)}\phi(x)\,,
 \)
so that the higher derivative term becomes
 \(
(\Box_{10}\phi)^2 = (\Box_{(d+1)}\phi)^2\,.
 \)
Substituting these results into the higher derivative part of the ten dimensional action, one obtains
\begin{equation}
S^{(10)}_{\rm HD} \supset \frac{\alpha^{\prime}}{2\kappa_{10}^2}\int d^{10}x\,\sqrt{-g_{10}}\;\left[1 + \gamma_1\,\alpha^{\prime} R_{10} + \gamma_2\,\alpha^{\prime} H_{MNP}H^{MNP} + \cdots \right](\Box_{10}\phi)^2\,.
\end{equation}

As the integrand is independent of the internal coordinates (apart from the metric and flux factors), we can integrate over the internal manifold \(M\) to obtain
\begin{eqnarray}
S_{\rm HD}^{(d+1)} \supset \frac{\alpha^{\prime}}{2\kappa_{10}^2}\int d^{d+1}x\,\sqrt{-g_{(d+1)}(x)}\;(\Box_{(d+1)}\phi)^2 \nonumber \\
\times \int_M d^{9-d}y\,\sqrt{g_M(y)} \left[1 + \gamma_1\,\alpha^{\prime} R_M + \gamma_2\,\alpha^{\prime} H_{mnp}H^{mnp} + \cdots \right]\,,
\end{eqnarray}
where in the reduced expression the ten dimensional curvature \(R_{10}\) and flux \(H_{MNP}H^{MNP}\) have been replaced by their internal components, \(R_M\) and \(H_{mnp}H^{mnp}\), respectively (here indices \(m,n,p\) run over the internal directions).

 It is then natural to define the effective higher derivative coupling in \((d+1)\) dimensions as
\begin{equation}
b = \frac{\alpha^{\prime}}{2\kappa_{10}^2}\int_M d^{9-d}y\,\sqrt{g_M(y)} \,\left[1 + \gamma_1\,\alpha^{\prime} R_M + \gamma_2\,\alpha^{\prime} H_{mnp}H^{mnp} + \cdots \right]\, = \alpha^{\prime} \, \mathcal{C}\,.
\end{equation}
 Hence, dimensional reduction compresses the full tower of stringy corrections into a single higher–derivative coupling
 \(
  \mathcal{C} \;=\; \frac{V_{M}}{2\kappa_{10}^{2}}
  \Bigl[
    1 - \tfrac14\,\alpha'\,\langle R_{M}\rangle
      + \tfrac1{24}\,\alpha'\,\langle H_{mnp}H^{mnp}\rangle
      + \dots
  \Bigr],
 \)
where $\gamma_1 =-\frac{1}{4}, \gamma_2 = \frac{1}{24}$,  $V_{M}$ is the internal volume, $\langle R_{M}\rangle$ the average internal curvature, and 
 $\langle H_{mnp}H^{mnp}\rangle$ the average flux density \cite{5a,GrossSloan1987}.
Their competition is captured by the dimensionless ratio
 \(
  \Xi \;\equiv\; \frac{6\langle H_{mnp}H^{mnp}\rangle}{\langle R_{M}\rangle}-1 ,
\)
so that $\Xi>0$ (flux dominance) yields $b>0$, whereas $\Xi<0$
(curvature dominance) gives $b<0$.
This expression represents the most general form for the effective coupling of the higher derivative term in the reduced theory. It explicitly shows that in addition to the contribution from the internal volume \(V_M\), there are corrections arising from the curvature of the internal manifold (via \(R_M\)) and from the background fluxes (via \(H_{mnp}H^{mnp}\)), as well as further higher order corrections represented by the ellipsis. These additional terms encapsulate the richer structure of the internal compactification, and they play a crucial role in determining the precise form of the generalized BF bound in the presence of $\alpha^{\prime}$ corrections.

Now as \(\mathcal{C}\) is the full \(\alpha^{\prime}\)-corrected coefficient that enters the higher-derivative coupling \(b=\mathcal{C}\alpha^{\prime}\), we can also define  
\(\mathcal{C}_{0}=V_{M}/(2\kappa_{10}^{2})\) is its tree-level value, fixed solely by the internal volume \(V_{M}\). So, we can define  
\(\delta\mathcal{C}=\mathcal{C}-\mathcal{C}_{0}\) as the correction generated by curvature and flux.  
Expanding \(R_{M}(y)\) and \(H_{mnp}H^{mnp}(y)\) in this basis isolates the zero-mode averages that control the sign and size of \(\delta\mathcal{C}\). In the important case where \(M\) is a Calabi–Yau three-fold, these zero-mode integrals are determined entirely by geometric data on the Calabi–Yau such as the (small) deviations from Ricci flatness and the distribution of background NS–NS flux quanta so that the resulting \(\delta\mathcal{C}\) is fixed by topological and flux numbers of the Calabi–Yau manifold \cite{Grana:2005jc}.
The sign and magnitude of the shift \(\delta\mathcal{C}=\mathcal{C}-\mathcal{C}_{0}\) can be understood by expanding the internal curvature and flux densities in a basis of orthonormal scalar harmonics \(Y_{I}(y)\) on \(M\).  Retaining only the zero–mode contributions \(R_{0}=\langle R_{M}\rangle\) and \(H_{0}=\langle H_{mnp}H^{mnp}\rangle\) one obtains, using the one-loop coefficients \(\gamma_{1}=-\tfrac14\) and \(\gamma_{2}=+\tfrac{1}{24}\) fixed in the ten-dimensional amplitude given by Gross-Sloan \cite{GrossSloan1987} and Metsaev-Tseytlin constructions \cite{Metsaev87}. Here, we have 
 \(
\frac{\delta\mathcal{C}}{\mathcal{C}_{0}}
  =-\frac{\alpha^{\prime}}{4}\,\langle R_{M}\rangle
   +\frac{\alpha^{\prime}}{24}\,\langle H_{mnp}H^{mnp}\rangle
   +\mathcal{O}\!\bigl(\alpha^{\prime 2}\langle\cdots\rangle\bigr).
 \)
It is convenient to package the curvature–flux competition in the dimensionless ratio
\(\Xi=6\langle H_{mnp}H^{mnp}\rangle/\langle R_{M}\rangle-1\); then \(\delta\mathcal{C}\) is positive, vanishing, or negative according as \(\Xi\) is greater than, equal to, or less than zero.  This single parameter therefore fixes, already at tree level in the reduction, whether the effective higher-derivative coupling \(b=\mathcal{C}\alpha^{\prime}\) is strengthened or suppressed once \(\alpha^{\prime}\) corrections are included.
For balanced non-Kähler vacua such as the Fu–Yau \(T^{2}\) fibration over \(K3\) \cite{Strominger86,Hull86,FuYau07,Becker2006} one finds nearly equal averages \(\langle R_{M}\rangle\simeq2.1\times10^{-2}\alpha^{\prime-1}\) and \(\langle H^{2}\rangle\simeq2.4\times10^{-2}\alpha^{\prime-1}\).  The corresponding \(\Xi\simeq0.14\) implies \(\delta\mathcal{C}/\mathcal{C}_{0}\approx+0.11\).  Gauge couplings in four dimensions, determined by \(g_{4}^{-2}=\mathcal{C}\), are strengthened, while the Kaluza–Klein scale \(M_{\mathrm{KK}}\propto\mathcal{C}^{-1/3}\) is reduced.  Such modest shifts are phenomenologically welcome because they allow unification near the conventional GUT scale without destabilising moduli.
Mirror half-flat manifolds arising from a single T-duality on a Calabi–Yau three-fold exhibit a richer pattern.  When the intrinsic torsion classes satisfy \(|\mathcal{W}_{1}|^{2}\simeq|\mathcal{W}_{2}|^{2}\simeq|\mathcal{W}_{3}|^{2}\simeq0.05\,\alpha^{\prime-1}\) \cite{ChiossiSalamon02,Gurrieri04}, one obtains \(\langle R_{M}\rangle\simeq1.0\times10^{-1}\alpha^{\prime-1}\) and \(\langle H^{2}\rangle\simeq1.5\times10^{-1}\alpha^{\prime-1}\).  If the phases align so that \(H\) tracks \(\mathcal{W}_{1}\) one gets \(\Xi\simeq0.80\) and \(\delta\mathcal{C}/\mathcal{C}_{0}\approx+0.07\); reversing the phase suppresses the effective \(\langle H^{2}\rangle\), giving \(\Xi\simeq-0.30\) and \(\delta\mathcal{C}/\mathcal{C}_{0}\approx-0.18\).  A discrete torsion flip therefore shifts the four-dimensional gauge coupling  and moves the BF stability bound by a comparable factor, highlighting the phenomenological leverage of half-flat flux choices.
Nearly-Kähler six-manifolds, exemplified by the round six-sphere \(S^{6}\) with radius \(R=5\sqrt{\alpha^{\prime}}\), carry larger positive curvature but comparatively small \(H\) flux.  Using \(R_{M}=5|\mathcal{W}_{1}^{+}|^{2}=5/(4R^{2})\) and \(H^{2}=1/(8R^{2})\) \cite{Lechtenfeld10} one finds \(\langle R_{M}\rangle=0.05\,\alpha^{\prime-1}\) and \(\langle H^{2}\rangle=0.0125\,\alpha^{\prime-1}\), yielding \(\Xi\simeq-0.75\).  This drives \(\delta\mathcal{C}/\mathcal{C}_{0}\approx-0.08\), lowers \(g_{4}^{-2}\), and raises \(M_{\mathrm{KK}}\).  The four-dimensional cosmological constant in the resulting AdS vacuum is \(\Lambda_{4}\simeq-1.6\times10^{-3}\alpha^{\prime-1}\); because \(m^{2}_{\mathrm{BF}}=\Lambda_{4}/3\), the negative shift in \(\mathcal{C}\) tightens the scalar stability window.
Nearly-parallel \(G_{2}\) manifolds interpolate between these extremes.  On the coset \(S^{7}=SO(8)/SO(7)\) with radius \(R=7\sqrt{\alpha^{\prime}}\) one has \(R_{M}=\tfrac{21}{8R^{2}}\approx0.026\,\alpha^{\prime-1}\).  With a \(G_{2}\)-instanton gauge bundle saturating the Bianchi identity, the flux average is \(\langle H^{2}\rangle\approx0.009\,\alpha^{\prime-1}\) \cite{Gemmer13}; the ratio \(\Xi\simeq+0.04\) leads to \(\delta\mathcal{C}/\mathcal{C}_{0}\approx+0.047\), an enhancement of the three-dimensional Planck mass, and a comparable tightening of the AdS\(_3\) BF bound despite the absence of supersymmetry.

Putting these examples together, one sees a clear trend: whenever positive curvature dominates, as in nearly-Kähler compactifications, \(\delta\mathcal{C}\) is negative; when fluxes compete effectively with or exceed curvature, as in balanced non-Kähler and some half-flat cases, \(\delta\mathcal{C}\) turns positive.  Now, as \(\mathcal{C}\) sets simultaneously the gauge coupling, the effective Planck mass, the Kaluza–Klein threshold, and the BF bound, a reliable assessment of the curvature–flux balance is indispensable for heterotic model building.  Mild positive shifts tend to aid gauge unification and reduce tachyonic risk, while large negative shifts can jeopardise moduli stability but may be exploited to raise the string scale.  Detailed discussion of solutions of the internal geometry thus serve not merely to close the ten-dimensional equations but to calibrate the entire hierarchy of scales in the \((d+1)\)-dimensional effective theory.

After integrating over the internal manifold \(M\), the effective action in \((d+1)\) dimensions for the scalar field \(\phi\), including the standard kinetic and mass terms along with the higher derivative correction, is given by (we have defined $g^{d+1}_{\mu\nu}$ as $g_{\mu\nu}$ for simplification)
\begin{equation}
S_{\rm eff} = \frac{1}{2}\int d^{d+1}x\,\sqrt{-g}\,\left[-g^{\mu\nu}\,\partial_\mu\phi\,\partial_\nu\phi - m^2\,\phi^2 - b\,(\Box\phi)^2\right]\,.
\end{equation}
The mass \(m^{2}\) is not a freely chosen parameter, but follows directly from the ten-dimensional dynamics already discussed.
First, the internal Laplacian acting on the zero–mode contributes the usual Kaluza–Klein term  
\(m_{\mathrm{KK}}^{2}=\lambda_{0}/V_{M}\),  
with \(\lambda_{0}\) the lowest scalar eigenvalue of the Laplacian on \(M\) and \(V_{M}\) its volume \cite{Candelas1985}.  
Second, the curvature– and flux–dressed operator \((\Box_{10}\phi)^{2}\) yields a shift  
\(\Delta m^{2}\propto\alpha^{\prime}\bigl(\gamma_{1}\langle R_{M}\rangle-\gamma_{2}\langle H_{mnp}H^{mnp}\rangle\bigr)\);  
this is precisely the curvature–flux competition encoded in the parameter \(\Xi\) that also governs the sign of the higher–derivative coupling \(b\) \cite{GrossSloan1987}. 
Finally, any flux or gauging required to satisfy the ten-dimensional Green–Schwarz–modified Bianchi identity supplies an additional term \(m_{\text{pot}}^{2}\) \cite{gc0}.  
Collecting the pieces, the reduced theory contains
\(
m^{2}=m_{\mathrm{KK}}^{2}+\Delta m^{2}+m_{\text{pot}}^{2},
 \)
so that both the magnitude of the scalar mass—are fixed by the same internal geometry and background flux data that determine \(b\) throughout the paper.
Varying the action with respect to \(\phi\) and performing the necessary integrations by parts leads to the modified equation of motion
\begin{equation}
\bigl(\Box - m^2 - b\,\Box^2\bigr)\phi = 0\,.
\end{equation}
The extra \(\Box^2\) term is a direct consequence of the higher derivative correction stemming from the internal compactification.
The effective action encapsulates the dynamics of a scalar field, including both the standard kinetic and mass contributions, as well as a higher derivative term. The higher derivative coupling \(b\), which is induced by the stringy $\alpha^{\prime}$ corrections, is determined by the detailed properties of the internal manifold -its volume, curvature, and background fluxes. This coupling ultimately modifies the equation of motion and plays a crucial role in understanding stringy corrections in lower dimensional theories. Moreover, the interplay between these corrections is central in determining the precise form of the generalized BF bound in the presence of $\alpha^{\prime}$ corrections.

 \section{Dual Conformal Field Theory}
We begin with the analysis of the effect of higher–derivative \(\alpha^{\prime}\) corrections on the spectrum and stability of a scalar field in AdS space, and we elucidate the corresponding implications for the dual conformal field theory (CFT) via the AdS/CFT correspondence. In the presence of such corrections, the standard   BF bound is modified to a generalized BF bound. This modification not only ensures bulk stability but also guarantees the unitarity of the dual CFT by preventing the appearance of complex operator dimensions.
Consider the Euclidean AdS\(_{d+1}\) metric
\begin{equation}\label{ds}
ds^2 = \frac{L^2}{z^2}\Bigl[dz^2 + \delta_{\mu\nu}\,dx^\mu dx^\nu\Bigr],
\end{equation}
where the radial coordinate \(z \to 0\) corresponds to the AdS boundary and the metric determinant is given by \(\sqrt{|g|}=(L/z)^{d+1}\). In this background, a scalar field is decomposed into Fourier modes along the boundary directions:
\begin{equation}
\phi(z,x)=\int\frac{d^dk}{(2\pi)^d}\,e^{i\,k\cdot x}\,f_k(z),
\end{equation}
with the near-boundary behavior of the radial profile assumed to be a power law,
\begin{equation}
f_k(z) \sim z^\beta.
\end{equation}
Using the d'Alembertian operator
\(
\Box = \frac{1}{\sqrt{g}}\,\partial_M\Bigl(\sqrt{g}\,g^{MN}\partial_N\Bigr),
 \)
one finds that the leading asymptotic behavior is
 \(
\Box\phi \simeq \frac{1}{L^2}\,\beta(\beta-d)\,e^{i\,k\cdot x}\,z^\beta,
 \)
while applying the operator twice yields
\( \Box^2\phi \simeq \frac{1}{L^4}\,[\beta(\beta-d)]^2\,e^{i\,k\cdot x}\,z^\beta.
 \)
 The modified linearized equation of motion, including the higher-derivative correction proportional to an effective coupling \(b\), is
\(
(\Box - m^{2} - b\,\Box^{2})\phi = 0 .
\)
Substituting the asymptotic expressions for \(\Box\phi\) and \(\Box^{2}\phi\) and multiplying by \(L^{4}\), we obtain
\(
L^{2}\,\beta(\beta-d) - m^{2}L^{4} - b\,[\beta(\beta-d)]^{2} = 0 .
\)
Now, after expanding and rearranging the terms, the equation can be recast as a quartic polynomial in \(\beta\)
\begin{eqnarray}
b\,\beta^{4} + 4b\,(1-d)\,\beta^{3}
+ \bigl(1 + b\,m^{2}L^{2} + 10b + 5bd + bd^{2}\bigr)\beta^{2} \nonumber \\[1mm]
+ \bigl(6b - d - 3bd + b\,d\,m^{2}L^{2}\bigr)\beta
- m^{2}L^{4} = 0 .
\label{betaroot}
\end{eqnarray}
In the limit \(b\to 0\) this quartic equation reduces to the familiar quadratic equation
\(
\beta(\beta-d)=m^2L^2,
 \)
which yields the canonical BF bound, \(m^2L^2\ge -d^2/4\). In the presence of the stringy corrections, however, the parameter \(b\) (which is of order \(\alpha^{\prime}\)) introduces a shift. Schematically, one may write
\begin{equation}
\beta = \beta_0 + \delta\beta,
\end{equation}
where \(\beta_0\) is the uncorrected solution and \(\delta\beta \propto \alpha^{\prime}\) encodes the higher–derivative (or \(1/N\)) correction.

According to the AdS/CFT correspondence, each bulk field is dual to an operator in the boundary CFT. In the standard two–derivative theory the asymptotic behavior of \(\phi\) determines the scaling dimension \(\Delta\) of the dual operator \(\mathcal{O}\) via
\begin{equation}
\Delta = \frac{d}{2} \pm \sqrt{\frac{d^2}{4}+m^2L^2}.
\end{equation}
When higher–derivative corrections are present, the relation between the asymptotic exponent \(\beta\) and the scaling dimension becomes modified. After an appropriate field redefinition (which can lead to \(\Delta = \beta\) or \(\Delta = d-\beta\), depending on the boundary conditions) the corrected scaling dimension is shifted as
\begin{equation}
\Delta = \Delta_0 + \delta\Delta,
\end{equation}
with \(\delta\Delta\) arising from \(\delta\beta\). The quartic equation (\ref{betaroot}) thus encodes the effect of stringy corrections on the operator dimensions. Unitarity of the CFT requires these dimensions to remain real; therefore, the generalized BF bound imposing that the quartic equation has only real solutions guarantees both the stability of the bulk and the consistency of the dual CFT.

More explicitly, one may express the corrected conformal dimension as
\begin{equation}
\Delta = \frac{d}{2} \pm \sqrt{\frac{d^2}{4}+m^2L^2+\delta(m^2L^2)},
\end{equation}
where \(\delta(m^2L^2)\) represents the perturbative correction induced by the higher–derivative term. These \(\alpha^{\prime}\) corrections, which appear in the effective coupling \(b\sim \mathcal{C}\,\alpha^{\prime}\), lead to finite–\(N\) (i.e. \(1/N\)) corrections to the dimensions of the dual operators.

 The AdS/CFT correspondence establishes a precise map between the asymptotic behavior of fields in the bulk and the conformal data of the dual CFT. In our treatment, the exponent \(\beta\) determines the scaling dimension \(\Delta\) of the operator \(\mathcal{O}\) dual to the bulk scalar field \(\phi\). When higher–derivative terms, arising naturally as \(\alpha^{\prime}\) corrections in string theory, are included in the bulk action, the standard second–order differential equation for \(\phi\) is replaced by a fourth–order equation. This change not only increases the richness of the solution space allowing for corrections of order \(\alpha^{\prime}\) but also directly impacts the dual CFT by modifying the operator dimensions.

The corrections encoded by \(b\sim \mathcal{C}\,\alpha^{\prime}\) are interpreted as \(1/N\) corrections in the dual CFT. Since unitarity in the CFT requires that operator dimensions be real, the condition that the quartic equation (\ref{betaroot}) have only real solutions is equivalent to demanding that the generalized BF bound holds. In practical terms, this analysis provides a quantitative probe into the subleading (finite–\(N\)) structure of the dual theory. It allows one to extract corrections to operator dimensions, correlation functions, and the overall phase structure of the CFT, offering deep insights into the quantum aspects of gravity and the full dynamical content of gauge/gravity duality.
Thus, the interplay between bulk stability and CFT unitarity mediated by the generalized BF bound serves as an essential consistency check for stringy corrections in holographic theories.

Now as the coupling that appears in the quartic equation \eqref{betaroot} is \(b=\mathcal{C}\alpha^{\prime}\), every vacuum fixes a definite value of \(\epsilon=b/L^{2}\) and hence a definite shift \(\delta\Delta\) in the dual operator dimensions.  
For definiteness let \(d=3\) so the uncorrected BF bound is \(m^{2}L^{2}\ge-9/4\) and take a scalar with \(m^{2}L^{2}=-2+\lambda\) where \(\lambda=0.05\) makes \(\Delta_{0}\simeq1.12\).
 The balanced geometry gives \(\delta\simeq+0.11\), hence \(\epsilon\approx8\times10^{-3}\).  
Equation \(\delta\Delta=-\epsilon(\Delta_{0}-3/2)(\Delta_{0}-1/2)/(1-2\Delta_{0}+3)\) then yields \(\delta\Delta\approx+3.0\times10^{-3}\).  
The dimension moves away from the unitarity edge, and the central charge changes by \(\delta c/c_{0}\approx-1.3\times10^{-3}\).
  When the torsion classes align with the flux one has \(\delta\simeq+0.07\rightarrow\epsilon\approx5\times10^{-3}\) and \(\delta\Delta\approx+1.9\times10^{-3}\).  If the phase is flipped, \(\delta\simeq-0.18\rightarrow\epsilon\approx-1.3\times10^{-2}\) and \(\delta\Delta\approx-5.0\times10^{-3}\).  
Thus a discrete torsion choice can shift \(\Delta\) by almost   in either direction and alter the BF stability margin by the same order.
  With \(\delta\simeq-0.08\) one finds \(\epsilon\approx-7\times10^{-3}\) and \(\delta\Delta\approx-2.7\times10^{-3}\).  
The generalized BF bound tightens; \(\lambda\) must exceed \(0.02\) to keep \(\Delta\) real.  
The central charge increases by \(+1.1\times10^{-3}\).
  The coset \(S^{7}\) gives \(\delta\simeq+0.047\rightarrow\epsilon\approx4\times10^{-3}\) and \(\delta\Delta\approx+1.5\times10^{-3}\).  
Although supersymmetry is absent, the positive shift pushes the spectrum further into the unitary regime.
These results exhibit a universal pattern: the sign of \(\Xi=6\langle H^{2}\rangle/\langle R_{M}\rangle-1\) dictates the sign of \(\delta\) and therefore whether operator dimensions rise (\(\Xi>0\)) or fall (\(\Xi<0\)).  The level changes in \(\delta\) induced by the internal geometry translate into level anomalous dimensions and central-charge shifts in the dual CFT, providing a clean, quantitative bridge between heterotic flux compactifications and finite-\(N\) observables on the boundary.

An advantageous approach to analyzing higher derivative theories is to introduce an auxiliary field $\chi$ defined by
\(
\chi \equiv \Box\phi\,.
\label{auxiliary}
\)
With this, the action  can be reformulated as
\begin{equation}
S_{\rm aux} = \frac{1}{2}\int d^{d+1}x\,\sqrt{-g}\,\Bigl[-g^{MN}\partial_M\phi\,\partial_N\phi - m^2\,\phi^2 + 2b\,\chi\,\Box\phi - b\,\chi^2\Bigr]\,.
\label{Saux}
\end{equation}
An integration by parts (and subsequent elimination of $\chi$) recovers the original action. This auxiliary field formulation reduces the higher derivative dynamics to coupled second order equations, which facilitates the extraction of the holographic RG flow and the evaluation of $\delta\Delta$ \cite{deBoer:2000,Skenderis:2002wp}. Introducing the auxiliary scalar \(\chi\) makes explicit that the higher–derivative term adds a new propagating degree of freedom whose mass is set by \(b^{-1}\sim(\alpha^{\prime}\mathcal{C})^{-1}\).  In the bulk this extra mode mixes with \(\phi\) through the off-diagonal coupling \(2b\,\chi\,\Box\phi\), and diagonalizing the quadratic form yields two second-order wave equations with shifted masses.  The lighter combination governs the near-boundary fall-off and therefore fixes the corrected conformal dimension \(\Delta_{0}+\delta\Delta\); the heavier combination decouples from the asymptotic region, encoding short-distance (\(z\ll L\)) stringy physics that manifests in the CFT as \(1/N\) contact-term corrections.  From an RG viewpoint the field \(\chi\) plays the role of a Pauli–Villars regulator that captures the finite-cutoff effects induced by \(\alpha^{\prime}\) corrections: integrating it out generates the quartic kinetic term for \(\phi\), while keeping it explicit allows one to track how these effects feed into running couplings and anomalous dimensions along the holographic radial direction.

A useful way to interpret this structure is through the holographic analogue of Lee–Wick regularization.  
Diagonalizing the quadratic form produces two bulk poles at \(p^{2}=-m_{-}^{2}\simeq-m^{2}\) and \(p^{2}=-m_{+}^{2}\simeq-1/b\).  
When \(b>0\) (flux-dominated compactifications) the heavy pole carries a negative residue and regulates ultraviolet behavior, precisely mirroring the Lee–Wick ghost of four-dimensional field theory \cite{Lee:1970iw,Grinstein:2008bg}. 
For \(b<0\) (curvature-dominated vacua) the rôle is reversed, but the generalized BF bound ensures the spectrum remains stable.  
In both cases the heavy mode resides at radial depth \(z\sim\sqrt{|b|}\ll L\) and does not generate an independent boundary primary; its imprint on the CFT appears only through the compactification-dependent shift \(\delta\Delta\) discussed above.  
Consequently, the single parameter \(\epsilon=b/L^{2}=\alpha^{\prime}\mathcal{C}/L^{2}\), fixed by the internal curvature–flux balance, simultaneously softens bulk amplitudes and dictates the size and sign of finite-\(N\) corrections in the dual conformal field theory, in agreement with the holographic renormalization  \cite{deBoer:2000,Skenderis:2002wp}.

Concrete evaluations of the curvature-flux ratio \(\Xi\) confirm its predictive power across several well-studied heterotic vacua.  
 In the balanced non-Kähler Fu-Yau \(T^{2}\!\rightarrow\!K3\) fibration  one has \(\langle R_{M}\rangle\simeq2.1\times10^{-2}\alpha^{\prime-1}\) and \(\langle H^{2}\rangle\simeq2.4\times10^{-2}\alpha^{\prime-1}\), giving \(\Xi\simeq0.14\) and a positive shift \(\delta\mathcal{C}/\mathcal{C}_{0}\approx+0.11\).  
 Mirror half-flat manifolds obtained by a single T-duality on a Calabi-Yau three-fold  yield \(\langle R_{M}\rangle\simeq1.0\times10^{-1}\alpha^{\prime-1}\) and \(\langle H^{2}\rangle\simeq1.5\times10^{-1}\alpha^{\prime-1}\); if the intrinsic-torsion phases align with the flux one finds \(\Xi\simeq0.80\) and \(\delta\mathcal{C}/\mathcal{C}_{0}\approx+0.07\), whereas reversing the phase drives \(\Xi\simeq-0.30\) and \(\delta\mathcal{C}/\mathcal{C}_{0}\approx-0.18\).  
 The nearly-Kähler six-sphere \(S^{6}\) of radius \(R=5\sqrt{\alpha^{\prime}}\) \cite{Lechtenfeld10} has \(\langle R_{M}\rangle=0.05\,\alpha^{\prime-1}\) and \(\langle H^{2}\rangle=0.0125\,\alpha^{\prime-1}\), giving \(\Xi\simeq-0.75\) and \(\delta\mathcal{C}/\mathcal{C}_{0}\approx-0.08\).  
 For the nearly-parallel \(G_{2}\) coset \(S^{7}=SO(8)/SO(7)\) with radius \(R=7\sqrt{\alpha^{\prime}}\) \cite{Gemmer13}, one finds \(\langle R_{M}\rangle\approx0.026\,\alpha^{\prime-1}\) and \(\langle H^{2}\rangle\approx0.009\,\alpha^{\prime-1}\), leading to \(\Xi\simeq+0.04\) and \(\delta\mathcal{C}/\mathcal{C}_{0}\approx+0.047\).  
 In every case, \(\Xi>0\) (flux dominance) correlates with a Lee–Wick pole that
softens bulk ultraviolet behavior, whereas \(\Xi<0\) (curvature dominance)
introduces a Pauli–Villars subtraction that remains compatible with the   generalized
BF bound.  Thus, these explicit examples show that the single geometric
invariant \(\Xi\) not only sets the magnitude of the higher–derivative
coupling \(b=\mathcal{C}\alpha^{\prime}\) but also determines whether the
heavy mode appears as a Lee–Wick pole (\(\Xi>0\)) or as a Pauli–Villars
regulator (\(\Xi<0\)), thereby governing the holographic regularization
mechanism across the heterotic landscape.

 Now we provide a detailed derivation of the modified scaling relations and renormalization group (RG) equations in the presence of $\alpha^{\prime}$ corrections.  
Depending on the boundary conditions, the conformal dimension $\Delta$ of the dual operator $\mathcal{O}$ is then identified as either $\Delta = \beta$ or $\Delta = d-\beta$.
Identifying the radial coordinate $z$ with the inverse energy scale $\mu\sim 1/z$, the radial evolution of $\phi$ implies \cite{deBoer:2000,Skenderis:2002wp}
\begin{equation}
z\frac{\partial\phi(z,x)}{\partial z} = \beta\,\phi(z,x)\,.
\label{RGflow1}
\end{equation}
In the dual CFT this maps onto the Callan Symanzik equation. In the two derivative theory the scaling behavior of the source $\phi_0(x)$ obeys
\begin{equation}
\mu\frac{\partial \phi_0}{\partial \mu} = -\Delta_0\,\phi_0\,,
\label{CSstandard}
\end{equation}
with $\Delta_0=\beta_0$ (or $d-\beta_0$).  
The inclusion of higher derivative terms modifies the  equation governing the near boundary behavior of the bulk field, leading to a shift $\beta = \beta_0 + \delta\beta$, and consequently altering the conformal dimension as $\Delta = \Delta_0 + \delta\Delta$.
So, with higher derivative corrections, the scaling dimension is shifted to
 \(
\Delta, \quad \delta\Delta\sim\delta\beta\,.
\label{Deltashift}
 \)
The RG flow equation is then modified to
\begin{equation}
\mu\frac{\partial \phi_0}{\partial \mu} = -\Bigl(\Delta_0+\delta\Delta\Bigr)\phi_0\,.
\label{CSmod}
\end{equation}
In a Wilsonian framework, the beta function for a coupling $g$ (dual to the operator $\mathcal{O}$) is given schematically by
\begin{equation}
\beta_{W}(g)=\mu\frac{\partial g}{\partial \mu}=\Bigl(\Delta_0+\delta\Delta\Bigr)g+\mathcal{O}(g^2)\,,
\label{beta_g}
\end{equation}
with $\delta\Delta$ incorporating finite coupling (or finite $N$) corrections resulting from the $\alpha^{\prime}$ effects.

Thus, the $\alpha^{\prime}$ corrections captured by the effective coupling $b$ provide finite coupling effects in the dual conformal field theory. Compactification fixes the sign and magnitude of the anomalous shift \(\delta\Delta\) through the single number \(\epsilon=b/L^{2}=\alpha^{\prime}\mathcal{C}/L^{2}\).  Because the linear term of the Wilsonian beta function is \(\beta_{W}(g)= (\Delta_{0}+\delta\Delta)\,g+\ldots\), any change in \(\epsilon\) translates directly into a change of slope for the RG flow.  In balanced Fu–Yau vacua the flux–curvature balance gives \(\epsilon\approx+8\times10^{-3}\), so \(\delta\Delta>0\) and the coefficient \(\Delta_{0}+\delta\Delta\) increases by roughly one per cent; an almost marginal coupling is driven away from criticality more quickly.  In mirror half-flat geometries the aligned-torsion branch still has \(\epsilon>0\) but smaller (\(\approx+5\times10^{-3}\)), yielding a gentler enhancement, while the phase-flipped branch flips the sign to \(\epsilon\approx-1.3\times10^{-2}\), reducing \(\Delta_{0}+\delta\Delta\) and flattening the flow.  Nearly-Kähler six-folds, dominated by curvature, give \(\epsilon\approx-7\times10^{-3}\) and further suppress \(\beta_{W}(g)\), whereas nearly-parallel \(G_{2}\) compactifications with \(\epsilon\approx+4\times10^{-3}\) provide a mild acceleration.  Thus positive \(\epsilon\) steepens the beta-function trajectory (couplings run faster), negative \(\epsilon\) flattens it (couplings run slower), and the internal curvature–flux balance in heterotic compactifications acts as a controllable dial for finite-\(N\) corrections in the boundary conformal field theory.

Physically, the compactification–induced shift \(\delta\Delta\) controls whether a nearly marginal operator becomes relevant or irrelevant once finite-\(N\) effects are included.  If \(\Delta_{0}<d\) (classically relevant) and \(\epsilon>0\), the larger \(\Delta_{0}+\delta\Delta\) drives the coupling \(g\) more steeply toward the infrared, potentially triggering symmetry breaking or a mass gap at a shorter RG time; the same positive \(\epsilon\) makes an irrelevant deformation with \(\Delta_{0}>d\) run faster to zero.  Conversely, \(\epsilon<0\) pushes a classically relevant operator closer to marginality, opening a “walking’’ window in which the theory lingers near scale invariance, while an irrelevant operator can move toward marginal relevance and seed a new infrared fixed point.  Thus Fu–Yau and aligned half-flat vacua (both \(\epsilon>0\)) accelerate flows, favoring rapid crossovers and early confinement, whereas torsion-flipped half-flat and nearly-Kähler vacua (\(\epsilon<0\)) slow the flow, allowing quasi-conformal behavior over several decades of scale.  Nearly-parallel \(G_{2}\) solutions sit in an intermediate regime, giving mild positive corrections that preserve qualitative features of the large-\(N\) fixed point but still generate calculable \(1/N\) shifts in critical exponents.  In all cases the curvature–flux balance encoded in \(\Xi\) acts as a geometric dial that selects between fast-running, walking, or fixed-point RG trajectories in the dual conformal field theory.

\section{Holographic Superconductors}
\noindent
Now building on the neutral–scalar analysis of the preceding section, we now {charge} the bulk field by introducing a minimal coupling to a background gauge field.   Our objective is to determine how  stringy corrections modify the critical temperature \(T_c\) at which the boundary theory undergoes superconducting condensation. 
In such holographic superconductors (see, e.g., \cite{Hartnoll2008a,Hartnoll2008b}), one considers a bulk action in the probe limit where the gravity sector is fixed and the matter fields (a gauge field and a charged scalar) evolve on this background. A typical bulk action is given by
\begin{equation}
S = \int d^{d+1}x\,\sqrt{-g}\left[ R + \frac{d(d-1)}{L^2} -\frac{1}{4}F_{MN}F^{MN} - |D\Psi|^2 - m^2|\Psi|^2 \right],
\end{equation}
where the covariant derivative is defined as
\begin{equation}
D_M = \nabla_M - i\,q\,A_M\,.
\end{equation}
Here, $\Psi$ is a charged complex scalar field and $F_{MN}$ is the field strength of the gauge field $A_M$. In the probe limit the backreaction of the matter fields on the metric is neglected, and the fixed background is typically an AdS black hole geometry.

Near the AdS boundary the charged scalar $\Psi$ admits an asymptotic expansion
 \(
\Psi(z,x) \sim z^{\Delta_-}\,\Psi_- + z^{\Delta_+}\,\Psi_+,
 \)
with the scaling dimensions defined by
 \(
\Delta_\pm = \frac{d}{2} \pm \sqrt{\frac{d^2}{4}+m^2L^2}\,.
\)
For spontaneous symmetry breaking it is standard to set the source $\Psi_- = 0$, so that the expectation value $\langle \mathcal{O} \rangle$ of the dual operator is given by $\Psi_+$.

Stringy (or $\alpha^{\prime}$) corrections introduce additional higher derivative contributions to the scalar sector. For instance, one might include an extra term of the form
\begin{equation}
S_{\rm HD} \supset \frac{b}{2}\int d^{d+1}x\,\sqrt{-g}\,(\Box\Psi)^2\,,
\end{equation}
with an effective coupling
 \(
b \sim \mathcal{C}\,\alpha^{\prime}\,,
 \)
where the  constant $\mathcal{C}$ encodes details of the internal compactification (such as its volume, curvature, and fluxes). These $\alpha^{\prime}$ corrections modify the effective mass of the scalar field, so that one must make the replacement
\begin{equation}
m^2 \to m^2+\delta(m^2)\,,
\end{equation}
where $\delta(m^2)$ is a perturbative correction induced by the higher derivative term.

As a consequence, the scaling dimension of the dual operator receives a corresponding shift. In the two derivative theory, the relation between the scalar mass and the scaling dimension which is given by
 \(
\Delta_0 = \frac{d}{2} \pm \sqrt{\frac{d^2}{4}+m^2L^2}\,.
 \)
gets modified, when  the correction is taken into account, and the corrected scaling dimension becomes
 \(
\Delta = \frac{d}{2} \pm \sqrt{\frac{d^2}{4}+m^2L^2+\delta(m^2L^2)}\,.
\)
Expanding for small $\delta(m^2L^2)$, one obtains
\begin{equation}
\Delta \approx \Delta_0 \pm \frac{1}{2\sqrt{\frac{d^2}{4}+m^2L^2}}\,\delta(m^2L^2)\,.
\end{equation}
Thus, even a small $\delta(m^2L^2)$ can lead to a significant fractional shift in $\Delta$ when the bare mass $m^2$ is near the BF bound (i.e., when $\sqrt{\frac{d^2}{4}+m^2L^2}$ is small).

The formation of a superconducting phase in the dual CFT is closely tied to the instability of the charged scalar field in the bulk. In particular, the scalar field condenses (developing scalar hair on the black hole) when the effective mass in the near-horizon region, which includes contributions from the gauge field (such as the negative term $-q^2 A_t^2$), falls below the IR BF bound. A negative shift in the effective mass (i.e., a negative $\delta(m^2)$) makes the scalar more tachyonic, thereby triggering the instability at a higher critical temperature $T_c$. In other words, the $\alpha^{\prime}$ corrections, by lowering the effective mass, tend to raise $T_c$.

Moreover, the condensate $\langle \mathcal{O} \rangle$, which is dual to the scalar field $\Psi$, is controlled by its scaling dimension. As $\Delta$ shifts toward the marginal value $\Delta\to d/2$, the condensate grows more rapidly as the temperature is lowered. In many models, this leads to a logarithmic or power-law divergence of $\langle \mathcal{O} \rangle$ in the probe limit when the mass approaches the BF bound. Therefore, the $\alpha^{\prime}$ corrections directly influence both the amplitude and the low-temperature behavior of the superconducting order parameter.

Since the coupling $b$ is of order $\alpha^{\prime}$ (or equivalently $1/\sqrt{\lambda}$ in the dual string theory), these corrections are interpreted as finite coupling or $1/N$ effects in the dual CFT. Consequently, operator dimensions, the critical temperature, and the scaling behavior of the condensate receive calculable subleading corrections beyond the leading large $N$ results.

Quantitatively, if we denote the shift in the scaling dimension by
\begin{equation}
\delta\Delta = \pm \frac{\delta(m^2L^2)}{2\sqrt{\frac{d^2}{4}+m^2L^2}}\,,
\end{equation}
one may estimate that the relative shift in the critical temperature is roughly
\begin{equation}
\frac{\Delta T_c}{T_c} \sim -\frac{\delta(m^2)}{|m^2|}\,.
\end{equation}
Now substituting the expression for $\delta m^2$ leads to
\begin{equation}
\frac{\Delta T_c}{T_c} \sim \mp \frac{2}{|m^2|L^2}\,\sqrt{\frac{d^2}{4}+m^2L^2}\,\delta\Delta\,.
\end{equation}
Thus, a more tachyonic effective mass (i.e., a negative $\delta(m^2)$) enhances the tendency for the scalar to condense by increasing $T_c$. Furthermore, the low-temperature scaling of $\langle \mathcal{O} \rangle$ becomes highly sensitive to $\Delta$, so that the changes in the effective scaling dimension directly affect the evolution and amplitude of the condensate.
The higher derivative $\alpha^{\prime}$ corrections, through the effective coupling $b\sim \mathcal{C}\,\alpha^{\prime}$, induce finite $N$ corrections in the dual CFT. These corrections manifest as shifts in the conformal dimension of the dual operator, leading to a  generalized BF bound. In the context of holographic superconductors, they lower the effective mass in the near-horizon region, raise the critical temperature $T_c$, and alter the low-temperature behavior of the condensate. Such effects provide a quantitative framework for understanding finite coupling corrections in strongly coupled field theories via the gauge/gravity duality.
 
The effective higher-derivative coupling that enters the superconductor action is fixed by the combination $b=\alpha^\prime\mathcal{C}$.  
Its sign and magnitude inherit all of the curvature–flux data through the shift $\delta\mathcal{C}=\mathcal{C}-\mathcal{C}_{0}$.  
Compactifications whose flux dominates over curvature (positive $\Xi$ and hence positive $\delta\mathcal{C}$) increase $b$ and make the near–horizon mass of $\Psi$ more negative.  
The scalar therefore becomes unstable at a higher temperature, so the critical temperature $T_{c}$ rises and the condensate forms more easily.  
Balanced non-Kähler Fu–Yau backgrounds provide a typical example: the modest positive shift $\delta\mathcal{C}/\mathcal{C}_{0}\!\simeq\!0.11$ strengthens the gauge coupling, lowers the effective mass, and raises $T_{c}$ without upsetting moduli stability.  
Mirror half-flat manifolds with aligned torsion give a similar but milder enhancement, whereas the phase–flipped branch, nearly-Kähler six–folds, and curvature–dominated $G_{2}$ spaces all yield negative $\delta\mathcal{C}$.  
In those cases $b$ is reduced, the scalar mass is driven upward, and superconductivity is suppressed: $T_{c}$ drops, the allowed $(m,L)$ window narrows, and the condensate grows more slowly.  
Thus the curvature–flux balance of the internal space selects whether the dual field theory realizes an easily triggered high–$T_{c}$ phase or a fragile low–$T_{c}$ phase, providing a direct link between stringy geometry and strongly coupled condensed-matter phenomena.

\section{Numerical Solution}

In principle, Eq.\,\eqref{betaroot} can be solved analytically for all four roots of the $\beta$ parameter. To present the results, we solve Eq.\,\eqref{betaroot} numerically for the smallest real root of $\beta$ (denoted $\beta_{1}$), which is of primary interest. The value of $\beta_{1}$ depends on the $b$, $m$, and $L$ parameters.

\begin{figure}[htb!]
    \centering
    \begin{tabular}{c c}
         \includegraphics[width=7cm,height=5.5cm]{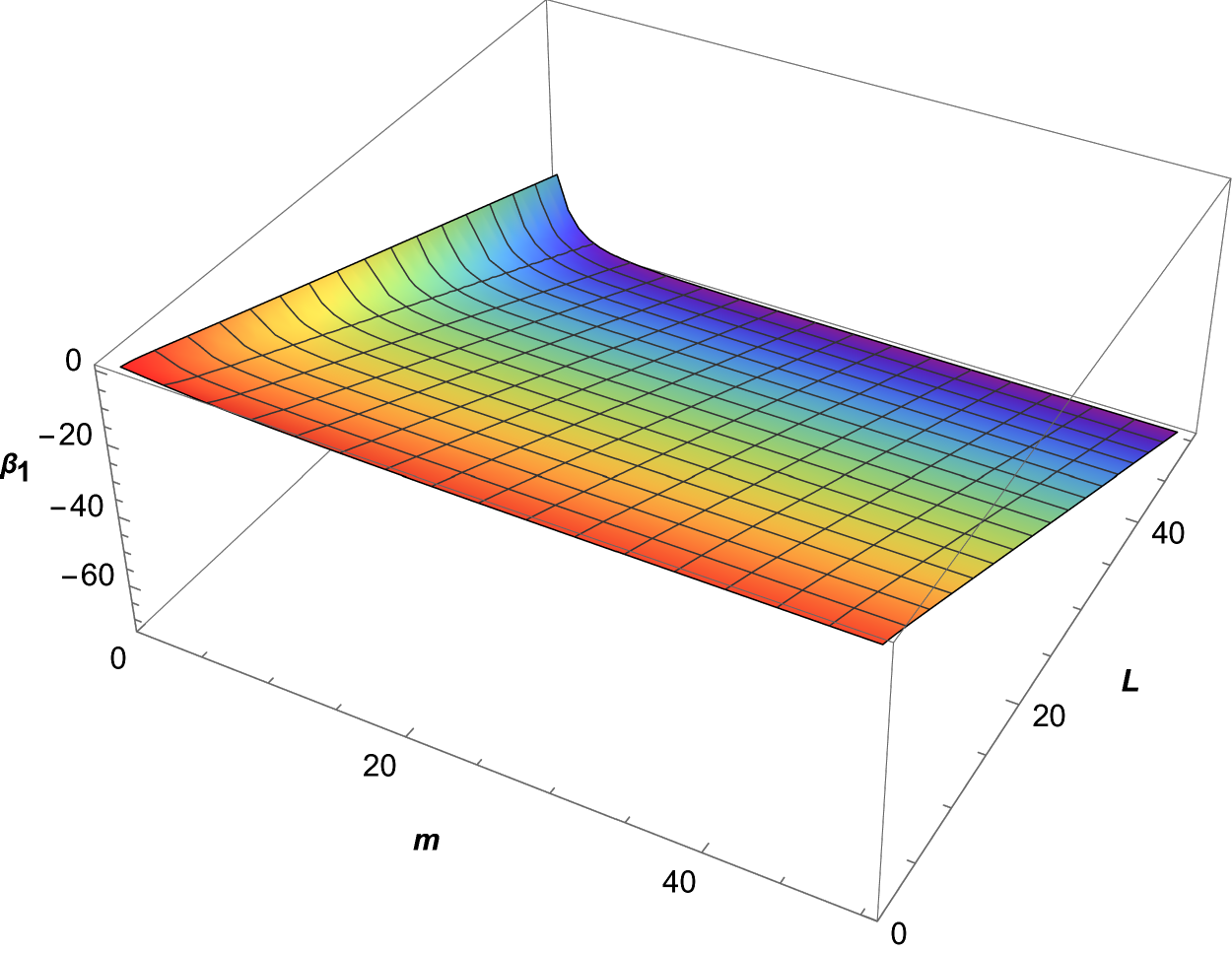} &
         \includegraphics[width=7cm,height=5.5cm]{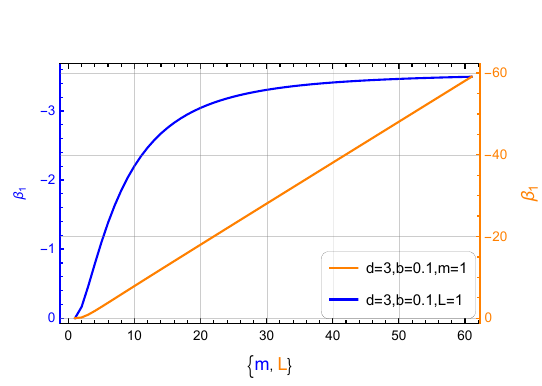} \\
         (a) & (b)
    \end{tabular}
    \caption{(Left): The smallest real root  plotted by varying the $m$ and $L$ parameters over the indicated range, while keeping $d=3$ and $b=0.1$. (Right): $\beta_{1}$ versus $m$ (blue curve) and $\beta_{1}$ versus $L$ (orange curve). Both curves share the same $x$-axis; the blue and orange labels indicate which variable is varied.}
    \label{xmlplot}
\end{figure}

\begin{figure}[htb!]
    \centering
    \includegraphics[scale=0.8]{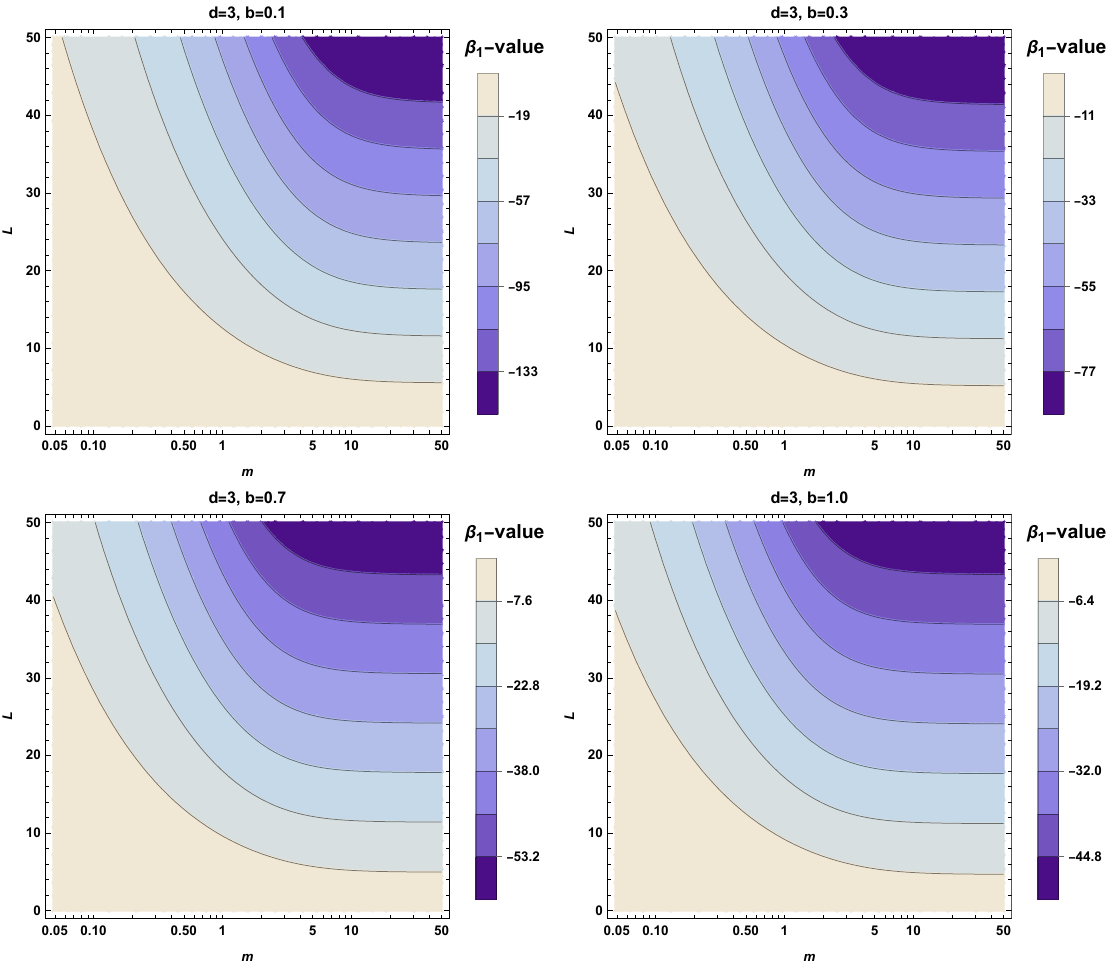}
    \caption{Contour plots showing the allowed parameter space in the $m$–$L$ plane for $d=3$ and $b\in\{0.1,0.3,0.7,1.0\}$. The colour bar indicates the value of $\beta_{1}$ along each contour.}
    \label{mld4}
\end{figure}

\begin{figure}[h!]
    \centering
    \includegraphics[scale=0.8]{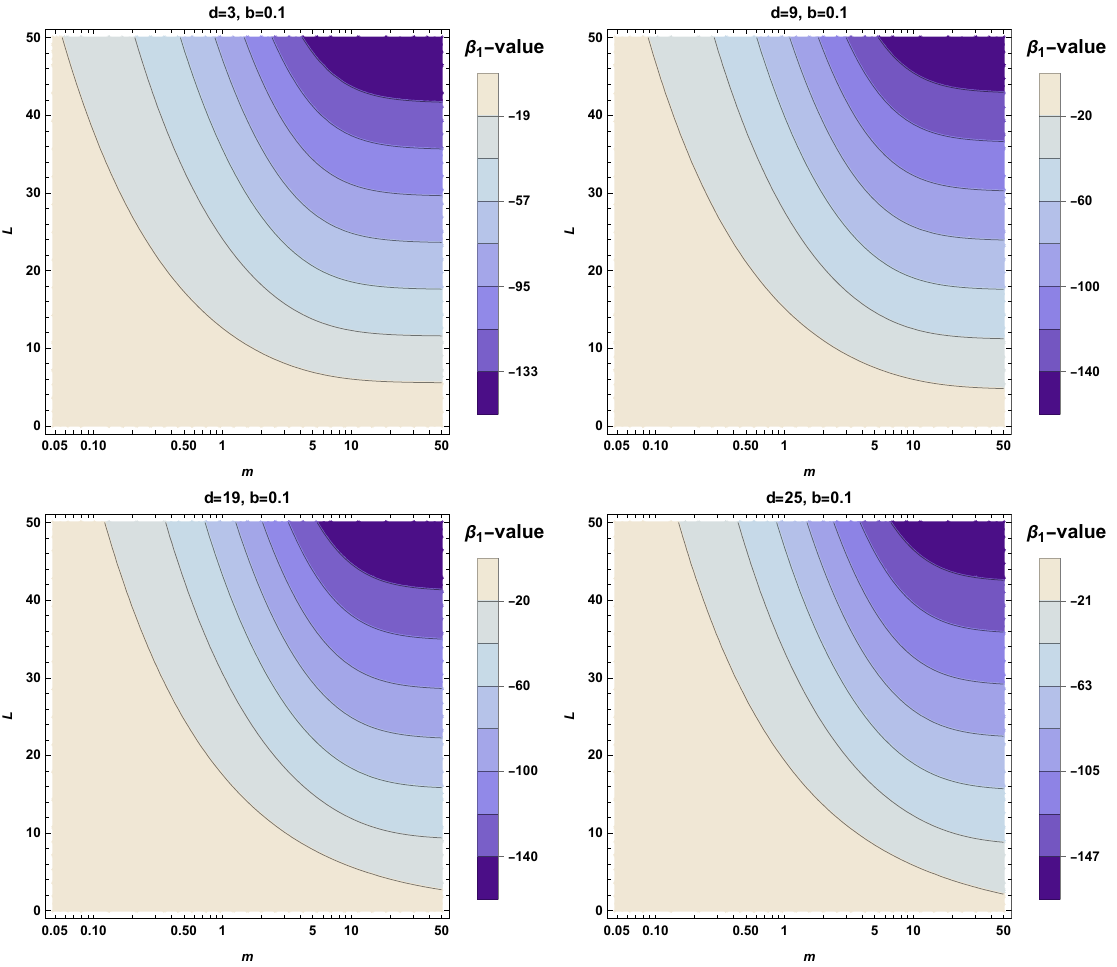}
    \caption{Contour plots showing the allowed parameter space in the $m$–$L$ plane for $b=0.1$ and $d\in\{3,9,19,25\}$. The colour bar indicates the value of $\beta_{1}$ along each contour.}
    \label{mldafix}
\end{figure}

\begin{figure}[h!]
    \centering
    \includegraphics[scale=0.8]{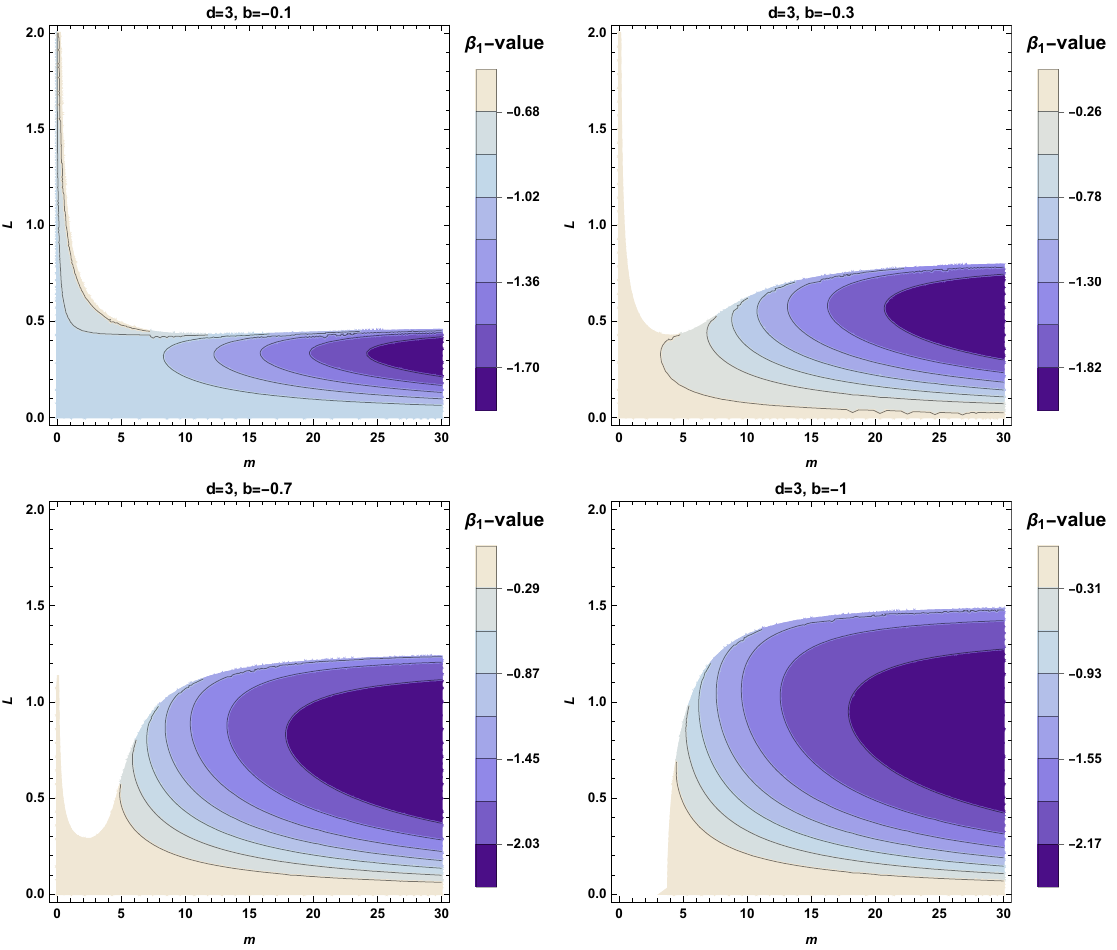}
    \caption{Contour plots showing the allowed parameter space in the $m$–$L$ plane for $d=3$ and negative values of $b\in\{-0.1,-0.3,-0.7,-1.0\}$. The colour bar indicates the value of $\beta_{1}$ along each contour.}
    \label{mlanegative}
\end{figure}

In Fig.\,\ref{xmlplot}(a) we plot the smallest real root of Eq.\,\eqref{betaroot} with $b=0.1$ and $d=3$, varying the mass ($m$) and length ($L$) parameters from $0$ to $50$. To understand how $\beta_{1}$ varies with $m$ and $L$, we plot $\beta_{1}$ against $m$ and $L$ in Fig.\,\ref{xmlplot}(b). For the orange curve we fix $b=0.1$, $d=3$, and $m=1$; from the plot we infer that $\beta_{1}$ is almost inversely proportional to the first power of $L$. The more interesting behaviour is the variation with respect to $m$: first $\beta_{1}$ decreases non-linearly as $m$ increases, and later it saturates for larger $m$. Thus the plot suggests an upper bound on the $m$ parameter.
In Fig.\,\ref{mld4} we plot contours in the $m$–$L$ plane after fixing $d=3$ and varying $b=0.1,\,0.3,\,0.7,\,1.0$. These contours represent the allowed parameter space obtained from the smallest real solution $\beta_{1}$ of Eq.\,\eqref{betaroot}. From Fig.\,\ref{mld4} it is clear that we cannot rule out any region of parameter space: all values of $m$ and $L$ are allowed. However, as we increase $b$, the values of $\beta_{1}$ increase, as indicated by the colour bars.
In Fig.\,\ref{mldafix} the allowed parameter space is plotted in the $m$–$L$ plane for fixed $b=0.1$ while varying $d$. The contours show that moving to higher dimensions produces only insignificant changes in the allowed parameter space.
Up to Fig.\,\ref{mldafix} there are no constraints on the allowed parameters: essentially all values of $m$ and $L$ are possible for a given $b$. However, for negative values of $b$ the allowed parameter space between $m$ and $L$ shrinks, as shown in Fig.\,\ref{mlanegative}. These contours provide lower and upper bounds on $m$ and $L$. For example, with $d=3$ and $b=-1.0$ (the fourth contour of Fig.\,\ref{mlanegative}) most values of $m$ and $L$ are ruled out for which $\beta_{1}$ is real; the allowed region is approximately $m\gtrsim5$ and $L\lesssim1.5$. All these results hint at bounds on the parameter space of the theory.

The numerical patterns in Figs.\,\ref{xmlplot}–\ref{mlanegative}
acquire a clear holographic meaning once we recall that the smallest real
root $\beta_{1}$ fixes the conformal dimension
$\Delta=\beta_{1}$ (or $\Delta=d-\beta_{1}$) of the dual scalar
operator, while the sign of the higher-derivative coupling
$b=\mathcal C\alpha^{\prime}$ tracks the balance between internal
flux and curvature.
For fixed bulk mass $m$, the approximate behaviour
$\beta_{1}\propto L^{-1}$ reflects that a larger AdS radius corresponds
to a larger central charge in the boundary theory, so operator dimensions
are suppressed as $N$ grows.
The saturation of $\beta_{1}$ at large $m$ marks the point where
the bare mass dominates over the $\alpha^{\prime}$ correction, setting an
effective upper bound on how far $\Delta$ can be lowered by mass
deformations alone.
Positive $b$ arises in flux-dominated compactifications; the associated
Lee–Wick pole carries a negative residue and regulates ultraviolet bulk
behaviour. Consequently, every $(m,L)$ pair leaves $\beta_{1}$ real,
so the generalized BF bound is automatically satisfied. Increasing $b$
pushes $\beta_{1}$ upward, rendering the dual operator more irrelevant
and reducing the danger of boundary instabilities.
Raising the boundary dimension produces only minor changes in the
allowed $(m,L)$ region: the canonical part of the equation rescales with
$d$, but the $\alpha^{\prime}$ term still governs the leading correction,
so the qualitative stability window is nearly dimension-independent for
moderate $b$.

Negative $b$ corresponds to curvature-dominated vacua. Here the
Lee–Wick pole acquires positive residue and would destabilize the theory
unless the generalized BF bound removes dangerous parameter space.
The shrinking wedges in the $(m,L)$ plane identify regions where the
heavy mode becomes tachyonic. For example, with $d=3$ and $b=-1$ the
theory remains unitary only for $m\gtrsim5$ and $L\lesssim1.5$.
Outside this island the operator dimension turns complex, and the dual
CFT loses unitarity.
Flux-controlled ($b>0$) compactifications therefore preserve stability
throughout the full $(m,L)$ range and merely renormalize operator
dimensions, whereas curvature-controlled ($b<0$) backgrounds admit
finite stability islands whose size quantifies proximity to a
Lee–Wick-type instability. These geometric constraints translate into
phenomenological bounds on Kaluza–Klein scales, condensate formation in
holographic superconductors, and the running of almost-marginal
couplings in the dual field theory.

\section{Conclusion}
 In this work we investigate how higher-derivative stringy \(\alpha^\prime\) corrections affect the low-energy effective action descended from heterotic string theory. Dimensional reduction compresses the entire tower of corrections into a single effective coupling that enters the \((d\!+\!1)\)-dimensional scalar sector. The resulting higher-order equation of motion yields a generalized BF bound, thereby altering field-stability criteria in AdS space. Now as the same coupling shifts operator dimensions, these bulk effects appear as finite-\(N\) corrections in the dual conformal field theory.
The coupling depends on the internal volume, the average curvature, and the background flux of the compactification manifold. The flux-dominated geometries relax the bound and widen the stable mass window, whereas curvature-dominated geometries tighten it and narrow that window. Simultaneously, the beta-function slope changes, causing renormalization-group flows to accelerate or decelerate. The heavy auxiliary mode generated by the higher-derivative term acts as a holographic Lee–Wick regulator.
When applied to holographic superconductors, the \(\alpha^\prime\) corrections shift the effective mass of the charged scalar near the horizon, modifying both the critical temperature and the low-temperature profile of the condensate. In particular, stringy corrections lower the effective bulk mass and raise $T_{c}$ when flux dominates, but raise the effective bulk mass and reduce $T_{c}$ when curvature dominates.
 Numerical analysis shows that positive values of the higher-derivative coupling accommodate a broad parameter range, whereas negative values impose stringent constraints. These findings highlight the important role of \(\alpha^\prime\) corrections in holography. 

This study provides a foundation for further research in this area and opens up several interesting problems for investigation. A detailed analytical treatment of the quartic equation governing the asymptotic behavior of the scalar field is needed to gain deeper insight into the corrections to the scaling dimensions. Extending the analysis beyond the probe approximation to include backreaction effects would provide a better understanding of the interplay between different backgrounds. It is also important to study the influence of higher-derivative corrections on other bulk fields, such as vector and spinor fields. This analysis can be used to develop a systematic framework that unifies these corrections across different sectors. Establishing a direct correspondence between the effective field theory approach and the results from string scattering amplitude computations could help determine the coefficients present in the higher-derivative expansion with greater precision. Further work in applying these theoretical insights to more realistic holographic models particularly those relevant to strongly coupled gauge theories or condensed matter systems is crucial. Refining our understanding of finite-coupling corrections and their effects on physical observables will be essential for deepening our knowledge of quantum many-body phenomena through gauge/gravity duality.

\bibliographystyle{ytphys.bst}

\end{document}